\documentclass[12pt]{article}
\usepackage{stmaryrd}
\usepackage{amssymb}
\usepackage{graphics}
\usepackage{epsfig}
\usepackage{a4wide}
\usepackage{cite}

\begin{document}

\title{Effects of Prediction Feedback in Multi-Route Intelligent Traffic Systems\footnote{Supported by National Natural Science Foundation
of China.}}
\author{ Chuanfei Dong$^\mathrm{a,c}$\footnote{dcfy@gatech.edu ,
dcfy@mail.ustc.edu.cn} , Xu Ma$^\mathrm{b,c}$ , Binghong
Wang$^\mathrm{c,d}$\footnote{bhwang@ustc.edu.cn ,
bhwangustc@hotmail.com} ,
\\
and Xiaoyan Sun$^\mathrm{c}$
\\
{$^\mathrm{a}$\small School of Earth and Atmospheric Sciences,
Georgia Institute of Technology,} \\
{\small Atlanta, GA 30332, U.S.A.}
\\
{$^\mathrm{b}$\small Department of Physics, Syracuse University,
Syracuse, NY 13244, U.S.A.}
\\
{$^\mathrm{c}$\small Department of Modern Physics and Nonlinear Science Center, University of }\\
{\small Science and Technology of China (USTC), Hefei, Anhui 230026,
P.R.China}
\\
{$^\mathrm{d}$\small The Research Center for Complex System Science,
University of Shanghai}\\
{\small for Science and Technology, Shanghai 200093, P.R.China} }

\date{}
\maketitle \vskip 15mm
\begin{abstract}
We first study the influence of an efficient feedback strategy named
prediction feedback strategy (PFS) based on a multi-route scenario
in which dynamic information can be generated and displayed on the
board to guide road users to make a choice. In this scenario, our
model incorporates the effects of adaptability into the cellular
automaton models of traffic flow. Simulation results adopting this
optimal information feedback strategy have demonstrated high
efficiency in controlling spatial distribution of traffic patterns
compared with the other three information feedback strategies, i.e.,
vehicle number and flux. At the end of this paper, we also discuss
in what situation PFS will become invalid in multi-route systems.

\end{abstract}


\maketitle

\vfill \eject

\baselineskip=0.32in

\renewcommand{\theequation}{\arabic{section}.\arabic{equation}}
\renewcommand{\thesection}{\Roman{section}.}
\newcommand{\nb}{\nonumber}

\newcommand{\Dir}{\kern -6.4pt\Big{/}}
\newcommand{\Dirin}{\kern -10.4pt\Big{/}\kern 4.4pt}
\newcommand{\DDir}{\kern -7.6pt\Big{/}}
\newcommand{\DGir}{\kern -6.0pt\Big{/}}

\makeatletter      
\@addtoreset{equation}{section}
\makeatother       

\section{Introduction}
\par
Vehicular traffic flow and related problems have triggered great
interests of a community of physicists in recent years because of
its various complex behaviors\cite{s1, s2, s3}. A lot of theories
have been proposed such as car-following theory\cite{s4}, kinetic
theory\cite{s5, s6, s7, s8, s9, s10, s11} and particle-hopping
theory\cite{s12, s13}. These theories have the advantages of
alleviating the traffic congestion and enhancing the capacity of
existing infrastructure. Although dynamics of traffic flow with
real-time traffic information have been extensively
investigated\cite{s14, s15, s16, s17, s18, s19}, finding a more
efficient feedback strategy is an overall task. Recently, some
real-time feedback strategies have been proposed, such as Travel
Time Feedback Strategy (TTFS)\cite{s14, s20}, Mean Velocity Feedback
Strategy (MVFS)\cite{s14,s21}, Congestion Coefficient Feedback
Strategy (CCFS)\cite{s14,s22}and Prediction Feedback Strategy
(PFS)\cite{s14,s23}. It has been proved that MVFS is more efficient
than that of TTFS which brings a lag effect to make it impossible to
provide the road users with real situation of each route\cite{s21}
and CCFS is more efficient than MVFS because random brake mechanism
of the Nagel-Schreckenberg(NS) model\cite{s12} brings fragile
stability of velocity\cite{s22}. However, CCFS is still not the best
one due to the fact that its feedback is not in time, so it cannot
reflect the real road situation immediately. Compared with CCFS, PFS
can provide road users with better guidance because it can predict
the future condition of the road. However, we never see these
advanced feedback strategies applied in a multi-route system in the
former work. In this paper, we first report the simulation results
adopting four different feedback strategies on a three-route
scenario with single route following the NS mechanism. We will also
discuss the situation of multi-route systems at the end of this
paper.

\par
The paper is arranged as follows: In Sec. II, the NS model and a
three-route scenario are briefly introduced, together with four
feedback strategies of TTFS, MVFS, CCFS and PFS all depicted in more
details. In Sec. III, some simulation results will be presented and
discussed based on the comparison of four different feedback
strategies. In the last section, we will make some conclusions.

\vskip 10mm
\section{THE MODEL AND FEEDBACK STRATEGIES}
\par
\textbf{A. NS mechanism}

\par
The Nagel-Schreckenberg (NS) model is so far the most popular and
simplest cellular automaton model in analyzing the traffic
flow\cite{s1,s2,s3,s12}, where the one-dimension CA with periodic
boundary conditions is used to investigate highway and urban
traffic. This model can reproduce the basic features of real traffic
like stop-and-go wave, phantom jams, and the phase transition on a
fundamental diagram. In this section, the NS mechanism will be
briefly introduced as a base of analysis.

\par
The road is subdivided into cells with a length of $\Delta$x=7.5 m.
Let \emph{N} be the total number of vehicles on a single route of
length \emph{L}, then the vehicle density is
$\rho$=\emph{N}/\emph{L}. $\emph{g}_{n}$(t) is defined to be the
number of empty sites in front of the \emph{n}th vehicle at time
\emph{t}, and $\emph{v}_{n}$(t) to be the speed of the \emph{n}th
vehicle, i.e., the number of sites that the \emph{n}th vehicle moves
during the time step \emph{t}. In the NS model, the maximum speed is
fixed to be $\emph{v}_{max}$=\emph{M}. In the present paper, we set
\emph{M}=3 for simplicity.

\par
The NS mechanism can be decomposed to the following four rules
(parallel dynamics):

\par
Rule 1. Acceleration: $\emph{v}_{i} \leftarrow
 min(\emph{v}_{i}+1,M)$;

\par
Rule 2. Deceleration: $\emph{v}_{i}^{'} \leftarrow
 min(\emph{v}_{i},\emph{g}_{i})$;

\par
Rule 3. Random brake: with a certain brake probability \emph{p} do
$\emph{v}_{i}^{''} \leftarrow max(\emph{v}_{i}^{'}-1,0)$; and

\par
Rule 4. Movement: $\emph{x}_{i} \leftarrow
\emph{x}_{i}+\emph{v}_{i}^{''}$;

\par
The fundamental diagram characterizes the basic properties of the NS
model which has two regimes called "free-flow" phase and "jammed"
phase. The critical density, basically depending on the random brake
probability \emph{p}, divides the fundamental diagram to these two
phases.

\par
\textbf{B. Three-route scenario}

\par
The three-route model, in which road users choose one of the three
routes according to the real-time information feedback, is similar
with the two-route model. The rules at the exit of the three-route
system, however, are more complex than that of the two-route system,
and we will explain the reason in Part C. In a three-route scenario,
it is supposed that there are three routes A, B and C of the same
length \emph{L}. At each time step, a new vehicle is generated at
the entrance of the system and will choose one route. If a vehicle
enters one of the three routes, its motion will follow the dynamics
of the NS model. As a remark, if a new vehicle is not able to enter
the desired route, it will be deleted. And a vehicle will also be
removed after it reaches the end point.

\par
Additionally, two types of vehicles are introduced: dynamic and
static vehicles. If a driver is a so-called dynamic one, he will
make a choice on the basis of the information feedback \cite{s20},
while a static one just enters a route at random ignoring any
advice. The density of dynamic and static travelers are
$\emph{S}_{dyn}$ and $1-\emph{S}_{dyn}$, respectively.

\par
The simulations are performed by the following steps: first, we set
the routes and board empty; second, after the vehicles enter the
routes, according to four different feedback strategies, information
will be generated, transmitted, and displayed on the board at each
time step. Finally, the dynamic road users will choose the route
with better condition according to the dynamic information at the
entrance of three routes.

\par
\textbf{C. Related definitions}

\par
The road conditions can be characterized by fluxes of three routes,
and the flux of one lane is defined as follows:
\begin{eqnarray}
\emph{F}=V_{mean}\rho=V_{mean}\frac{N}{L}
\end{eqnarray}
where $V_{mean}$ represents the mean velocity of all the vehicles on
one of the roads, \emph{N} denotes the vehicle number on each road,
and \emph{L} is the length of three routes. Then we describe four
different feedback strategies, respectively.

\par
TTFS: At the beginning, all routes are empty and the information of
travel time on the board is set to be the same. Each driver will
record the time when he enters one of the routes. Once a vehicle
leaves the three-route system, it will transmit its travel time on
the board and at that time a new dynamic driver will choose the road
with shortest time.

\par
MVFS: Every time step, each vehicle on the routes transmits its
velocity to the traffic control center which will deal with the
information and display the mean velocity of vehicles on each route
on the board. Road users at the entrance will choose one road with
largest mean velocity.

\par
CCFS: Every time step, each vehicle transmits its signal to
satellite, then the navigation system (GPS) will handle that
information and calculate the position of each vehicle which will be
transmitted to the traffic control center. The work of the traffic
control center is to compute the congestion coefficient of each road
and display it on the board. Road users at the entrance will choose
one road with smallest congestion coefficient.

\par
The congestion coefficient is defined as
\begin{equation}
C=\sum_{i=1}^{m} n_i^{w}.
\end{equation}
Here, $\emph{n}_{i}$ stands for vehicle number of the \emph{i}th
congestion cluster in which cars are close to each other without a
gap between any two of them. Every cluster is evaluated by a weight
\emph{w}, here \emph{w}=2\cite{s22}.

\par
PFS: It is based on CCFS because CCFS is the best one among the
three strategies above.

\par
Every time step, the traffic control center will receive data from
the navigation system (GPS) like CCFS. The work of the center is to
compute the congestion coefficient of each lane and simulate the
future road situation based on the current road situation adopting
the CCFS and display the prediction congestion coefficient on the
board. Road users at the entrance will choose one road with smallest
prediction congestion coefficient. For example, if the prediction
time ($\emph{T}_{p}$) is 50 seconds and the current time is 100th
second, the traffic control center will simulate the road situation
at the next 50 seconds using CCFS, predict the road situation at
150th second, and show the result on the board at the entrance of
the road. Finally the road users at 100th second will choose one
lane with smallest prediction congestion coefficient at 150th second
predicted by the new strategy. So as to analogize, the road user at
the entrance at 101th second will choose one road with smallest
prediction congestion coefficient at 151th second predicted by this
strategy as explained above and so on.

\par
In this paper, the three-route system has only one entrance and one
exit instead of one entrance and three exits as shown in Fig.1. So
the road condition in this paper is closer to the reality. The rules
at the exit of the three-route system are as follows:
\begin{figure}[htbp]
\vspace*{-0.3cm} \centering
\includegraphics[scale=0.7]{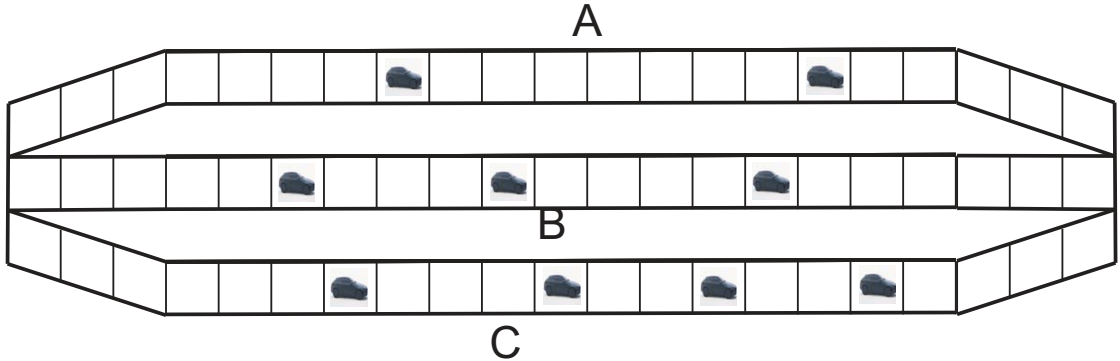}
\vspace*{-0.3cm} \centering \caption{\label{fig1} The three-route
system only has one entrance and one exit.}
\end{figure}

\par
(a) At the end of three routes, the nearest vehicle to the exit goes
first.

\par
(b) If the vehicles at the end of three routes have the same
distance to the exit, fastest one drives, first it goes out.

\par
(c) If the vehicles at the end of three routes have the same speed
and distance to the exit, the vehicle in the route which owns most
vehicles drives out first.

\par
(d) If the rules (a), (b) and (c) are satisfied at the same time,
then the vehicles go out randomly.

\par
Though the rules in the three-route system seem to be the same as
that in the two-route system, if you consider it carefully, you may
find out that rules in the three-route system are much more complex.
For example, among the vehicles in route A, B and C, the vehicle on
route A is the nearest one to the exit and meanwhile the vehicles on
route B and C have the same distance to the exit which will never
happen in the two-route system. In the following section,
performance by using four different feedback strategies will be
shown and discussed in more details.

\vskip 10mm
\section{SIMULATION RESULTS}

\par
All simulation results shown here are obtained by 90000 iterations
excluding the initial 5000 time steps. From the data shown above, we
can find out that the time steps needed to reach stable state in a
three-route system are much longer than that in a two-route system,
where it only needs 25000 time steps to reach stable
state\cite{s23}. So it brings about a lot of difficulties in our
current work. Fig.2 shows the dependence of average flux and
prediction time($\emph{T}_{p}$) adopting prediction feedback
strategy. As to the routes' processing capacity, we can see that in
Fig.2, the prediction time($\emph{T}_{p}$) corresponding to the
highest value of the average flux is about 260 time steps which are
much longer than that before\cite{s23}. Hence, we will use
$\emph{T}_{p}$=260 in the following paragraphs.
\begin{figure}
\centering
\includegraphics[scale=1.0]{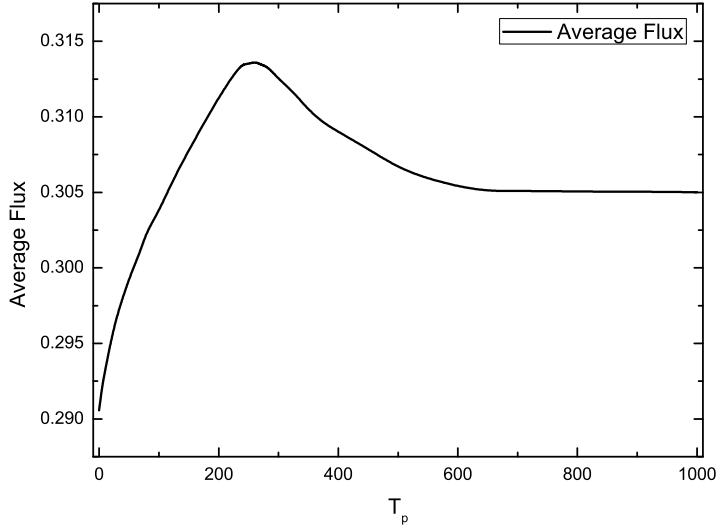}
\caption{\label{fig2} Average flux \emph{vs} prediction
time($\emph{T}_{p}$). The parameters are \emph{L}=2000,
\emph{p}=0.25, and $\emph{S}_{dyn}$=0.5.}
\end{figure}

\par
In contrast with PFS, the fluxes of three routes adopting CCFS, MVFS
and TTFS show oscillation (see Fig.3) obviously due to the
information lag effect\cite{s22}. This lag effect can be understood
because the other three strategies cannot reflect the road current
conditions. For TTFS, the travel time reported by a driver at the
end of three routes only represents the road condition in front of
him, and perhaps the vehicles behind him have got into the jammed
state. Unfortunately, this information will induce more vehicles to
choose his route until a vehicle from the jammed cluster leaves the
system. This effect apparently does harm to the system. For MVFS, we
have mentioned that the NS model has a random brake scenario which
causes the fragile stability of velocity, so MVFS cannot completely
reflect the real condition of routes. The other reason for the
disadvantage of MVFS is that flux consists of two parts, mean
velocity and vehicle density, but MVFS only grasps one part and
lacks the other part of flux.  Another reason for the oscillation of
three former strategies is that the three-route system only has one
exit, therefore, only one car can go out at each time step, which
may result in the traffic jams to happen at the end of the routes.
However, the new strategy can predict the effects on the route
condition caused by the traffic jams happened at the end of the
routes, then try to avoid the traffic jams happening to the best of
its ability and alleviate the negative effects as much as possible.
Here, we want to stress that though PFS try to avoid the jammed
state, the structure of the traffic system (one exit) still make
jams possible to happen at the end of the route occasionally, also
this can explain the slight oscillation in Fig.3(d). Hence, the new
strategy may greatly improve the road situation. Compared with CCFS,
the performance adopting PFS is remarkably improved, not only on the
value but also the stability of the flux. Hence, as to the flux of
the three-route system, PFS is the best one.
\begin{figure}
\centering
\includegraphics[scale=0.7]{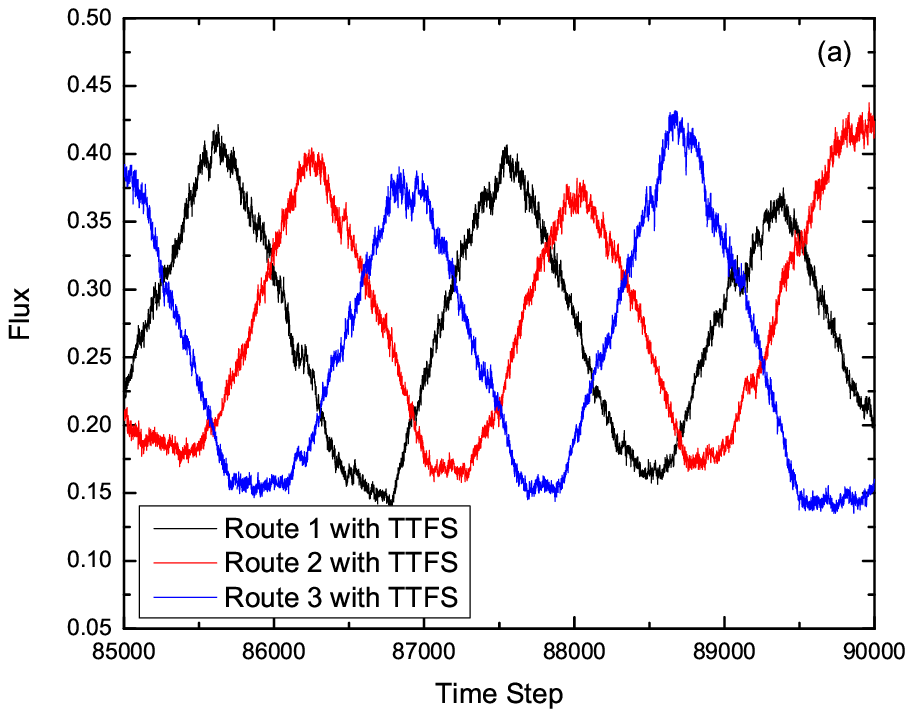}
\includegraphics[scale=0.7]{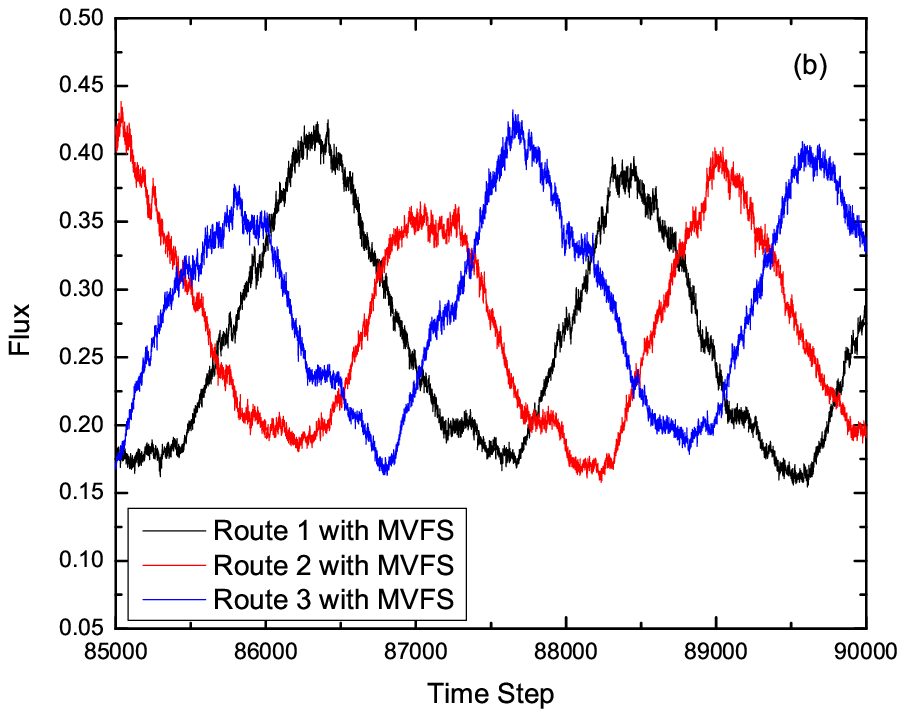}
\includegraphics[scale=0.7]{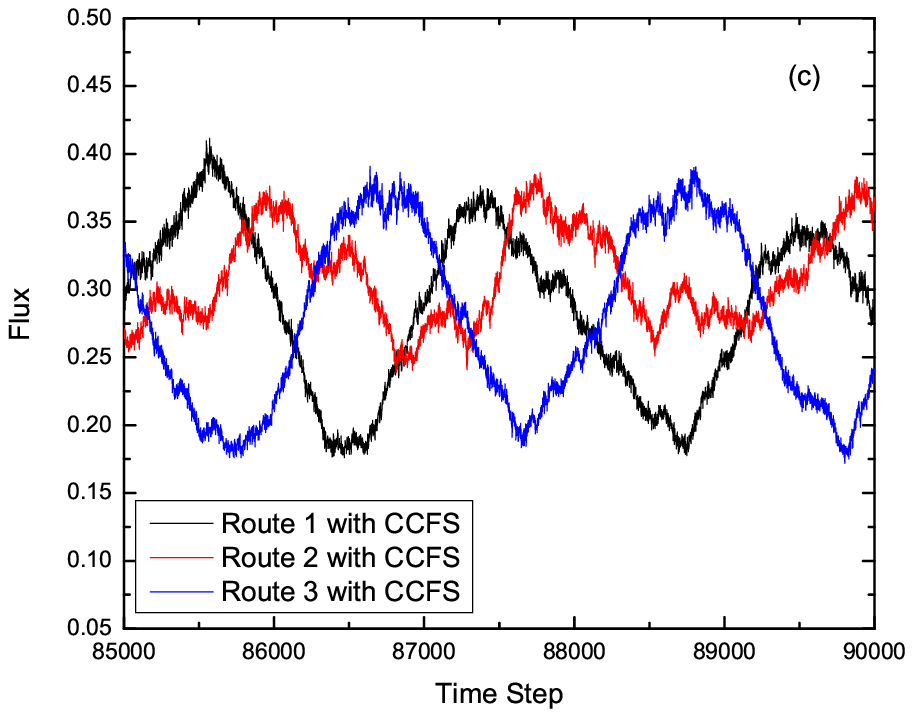}
\includegraphics[scale=0.7]{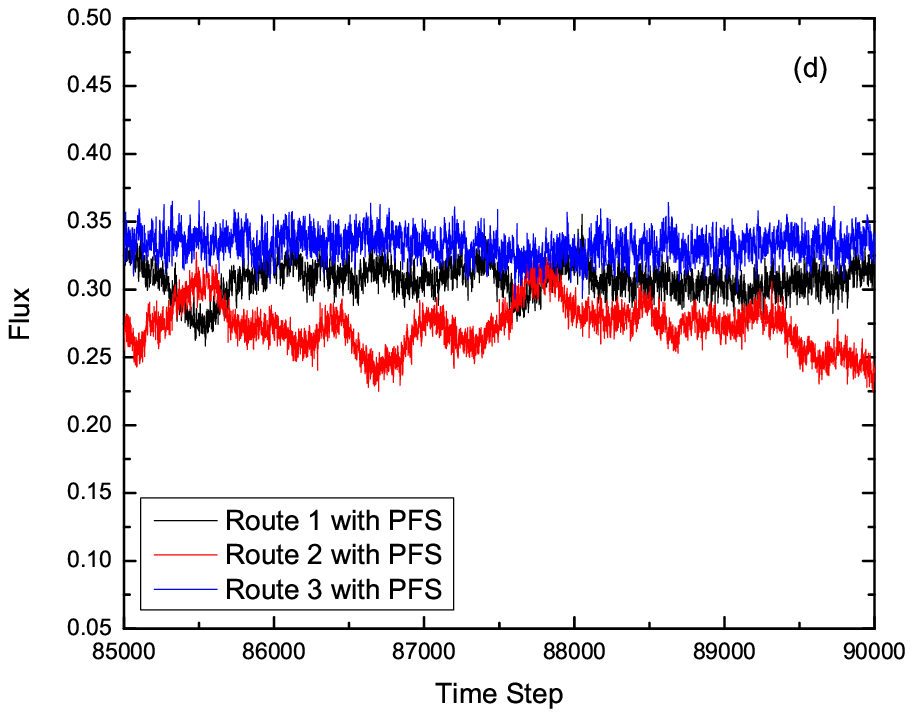}
\caption{\label{fig3} (Color online)(a) Flux of each route with
travel time feedback strategy. (b) Flux of each route with mean
velocity feedback strategy. (c) Flux of each route with congestion
coefficient feedback strategy. (d) Flux of each route with
prediction feedback strategy. The parameters are \emph{L}=2000,
\emph{p}=0.25, $\emph{S}_{dyn}$=0.5, and $\emph{T}_{p}$=260. }
\end{figure}

\par
In Fig.4, vehicle number versus time step shows almost the same
tendency as Fig.3, and that the routes' accommodating capacity is
greatly enhanced with an increase in average vehicle number from 230
to 870, so perhaps the high fluxes of three routes with PFS are
mainly due to the increase of vehicle number. Here, someone may ask
if the high vehicle number will result in the jammed cluster in each
route. As to the routes' stability, we know PFS is the best one (see
Fig.3 and Fig.4), which means the vehicles should be almost
uniformly distributed on each route instead of being together at the
end of the routes. Furthermore, even there are 870 vehicles on each
route, most vehicles can still occupy 2 sites on each lane and a few
vehicles may even occupy 3 route sites because the total length of
each route is fixed to be 2000 sites. Meanwhile, this means there is
only one site between most of the vehicles. Though the vehicles are
almost distributed separately on each lane, the one exit structure
make jams still have a chance to happen at the end of the route,
however, PFS can prevent jams from further expanding and alleviate
the negative effects as much as possible, so that the jammed state
will disappear soon. So as to analogize, even the jams happen again,
the poor road condition will be relieved in a short time. Hence,
there is some connection between the high accommodating capacity
shown in Fig.4 and the traffic jams discussed in the paragraph
above.
\begin{figure}
\centering
\includegraphics[scale=0.7]{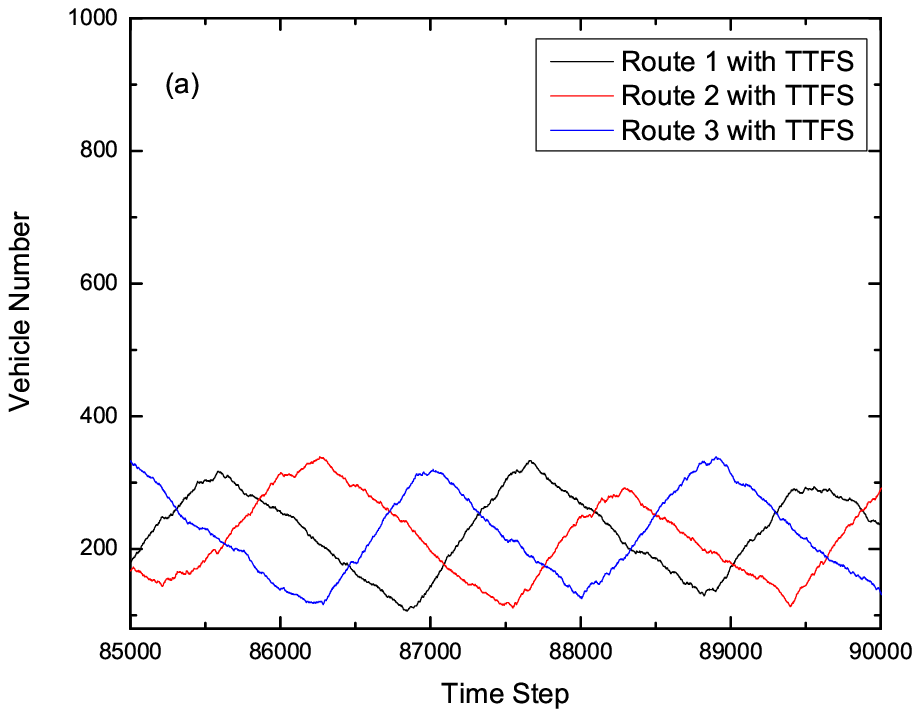}
\includegraphics[scale=0.7]{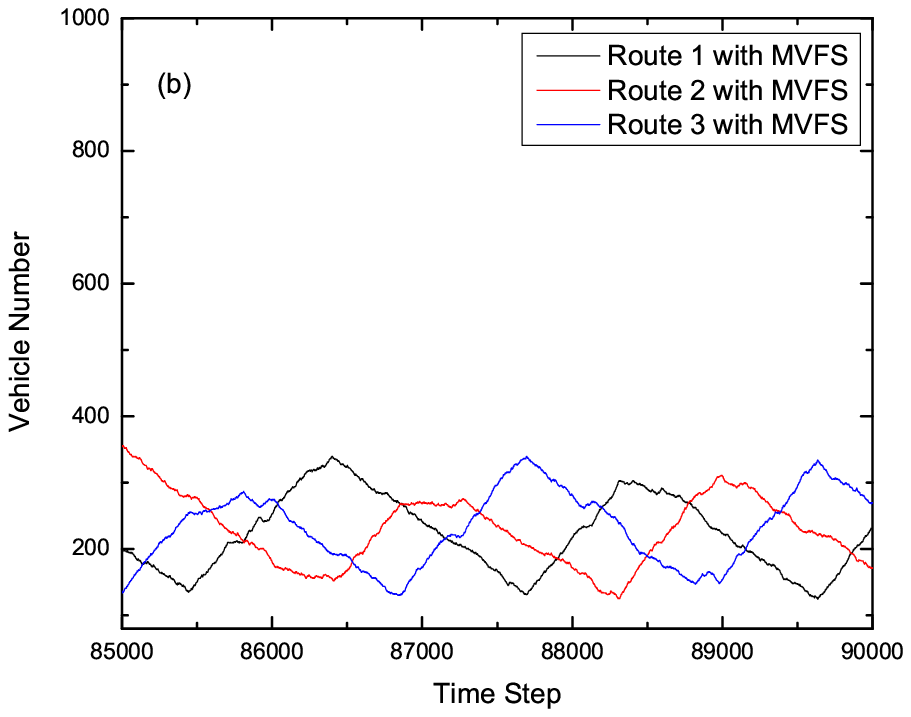}
\includegraphics[scale=0.7]{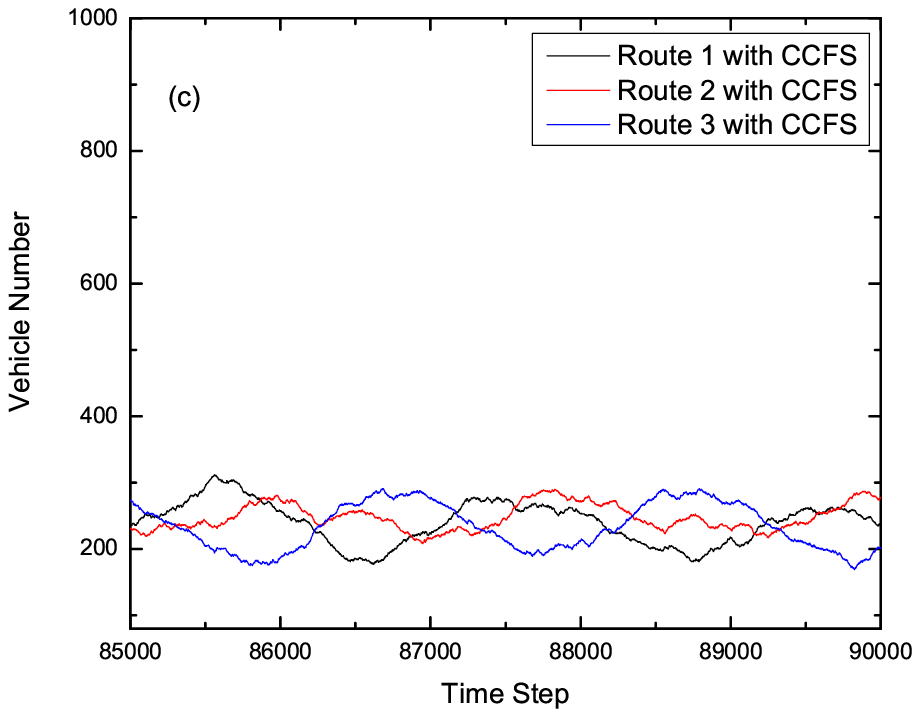}
\includegraphics[scale=0.7]{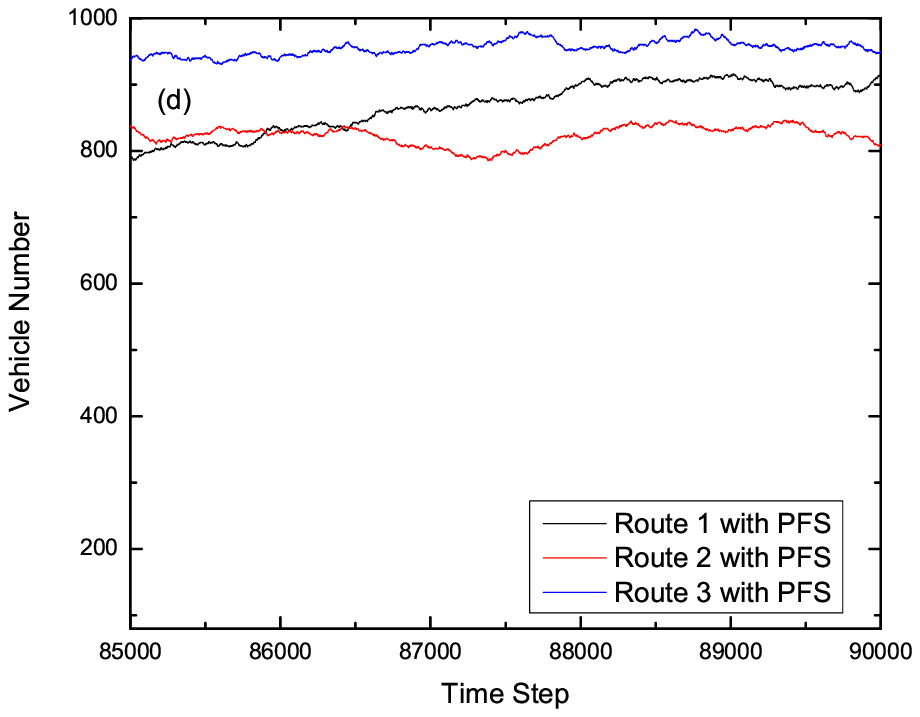}
\caption{\label{fig4} (Color online)(a) Vehicle number of each route
with travel time feedback strategy. (b) Vehicle number of each route
with mean velocity feedback strategy. (c) Vehicle number of each
route with congestion coefficient feedback strategy. (d) Vehicle
number of each route with prediction feedback strategy. The
parameters are set the same as in Figure 3.}
\end{figure}

\par
In Fig.5, speed versus time step shows that although the speed is
the stablest by using prediction feedback strategy, it is the lowest
among the four different strategies. The reason is that the speed
partially depends on the number of empty sites between two vehicles
on the lane. As mentioned before, the routes' accommodating capacity
is the best by using PFS, indicating the speed adopting PFS the
lowest. From the stability of the velocity, we infer that the
vehicles should drive at almost uniform speeds on each route.
Without consider other factors, the speed should be a little more
than one because there is only one site between most of the vehicles
as mentioned above and the vehicle behind another vehicle can move
at most the current empty sites between them which is also required
by NS mechanism. If we take the random brake effects and the
occasional jams at the end of the route into account, the vehicles'
average velocity low than one is possible and reasonable; therefore,
the average velocity $V_{avg}\sim 0.7$ in this paper could be
understood and there should be no conflicts between the average
velocity and the almost uniform distribution of vehicles on each
route. Other three strategies' high speeds may result from the
vehicles at the beginning or middle parts of the lane which have
high velocities instead of the vehicles near the exit, because the
average velocity in each lane depends on all vehicles' velocities.
Fortunately, flux consists of two parts, mean velocity and vehicle
density. Therefore, as long as the vehicle number (because the
vehicle density is $\rho$=\emph{N}/\emph{L}, and the \emph{L} is
fixed to be 2000, so $\rho$ $\propto$ vehicle number (\emph{N})) is
large enough, the flux can also be the largest.
\begin{figure}
\centering
\includegraphics[scale=0.7]{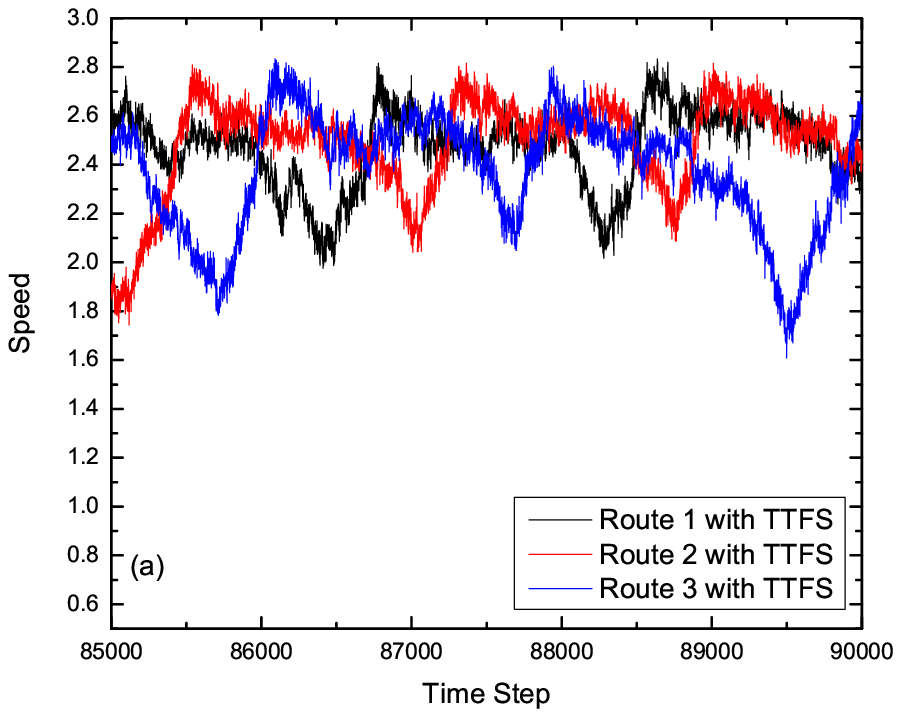}
\includegraphics[scale=0.7]{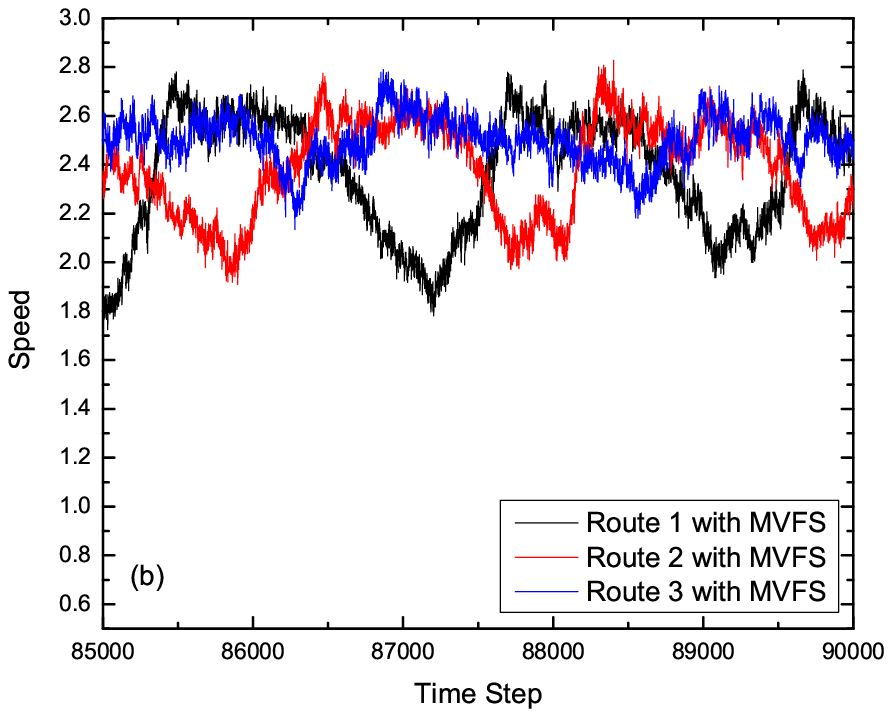}
\includegraphics[scale=0.7]{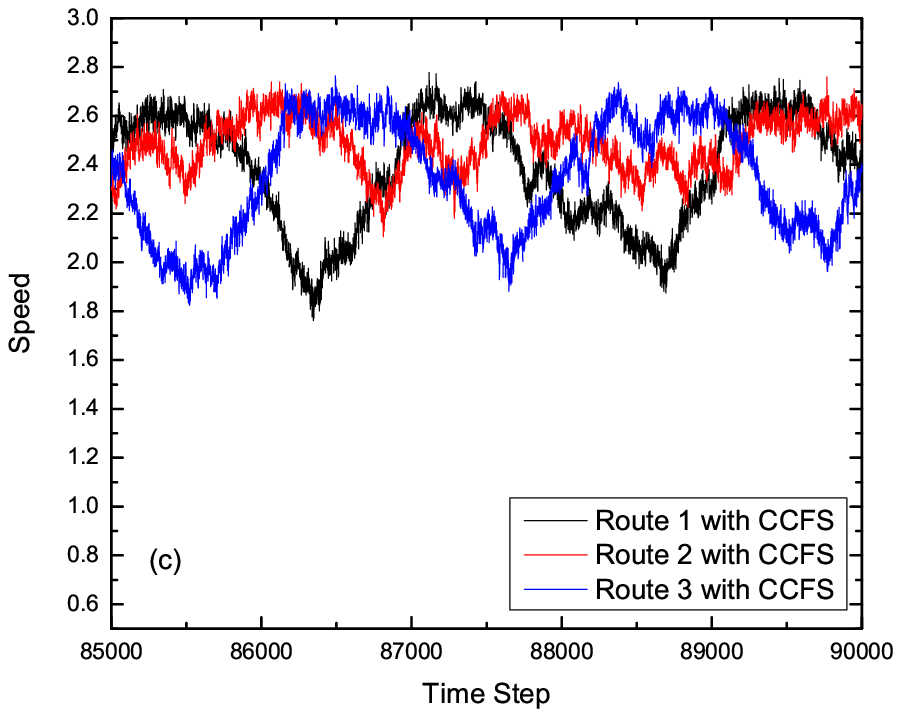}
\includegraphics[scale=0.7]{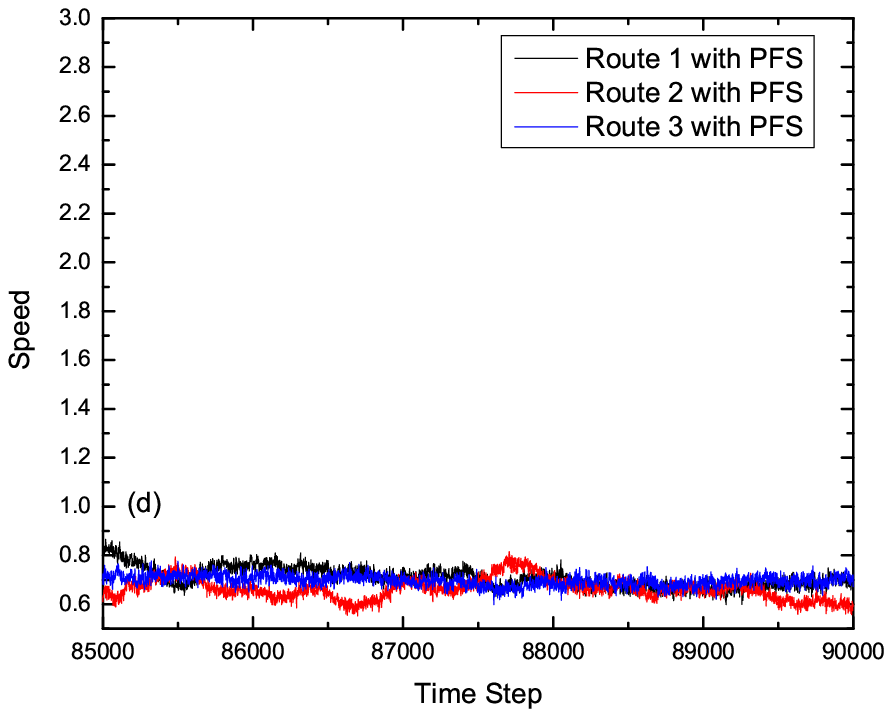}
\caption{\label{fig5} (Color online)(a) Average speed of each route
with travel time feedback strategy. (b) Average speed of each route
with mean velocity feedback strategy. (c) Average speed of each
route with congestion coefficient feedback strategy. (d) Average
speed of each route with prediction feedback strategy. The
parameters are set the same as in Figure 3.}
\end{figure}

\par
Fig.6 shows that the average flux fluctuates feebly with a
persisting increase of dynamic travelers by using four different
strategies. As to the routes' processing capacity, the prediction
feedback strategy is proved to be the best one because the flux is
always the largest at each $\emph{S}_{dyn}$ value and enhances with
a persisting increase of dynamic travelers. A question here for some
readers is why the average fluxes in Fig.6 adopting four different
strategies are smaller than that (see Fig.7) shown in the former
work\cite{s23} where the traffic system has only two routes and the
prediction time($\emph{T}_{p}$) is fixed to be 60. The reason is
that the three-route system in this paper still permits at most one
car to enter the entrance at each time step. Hence, more routes the
traffic system owns, lower average fluxes it has.
\begin{figure}
\centering
\includegraphics[scale=1.2]{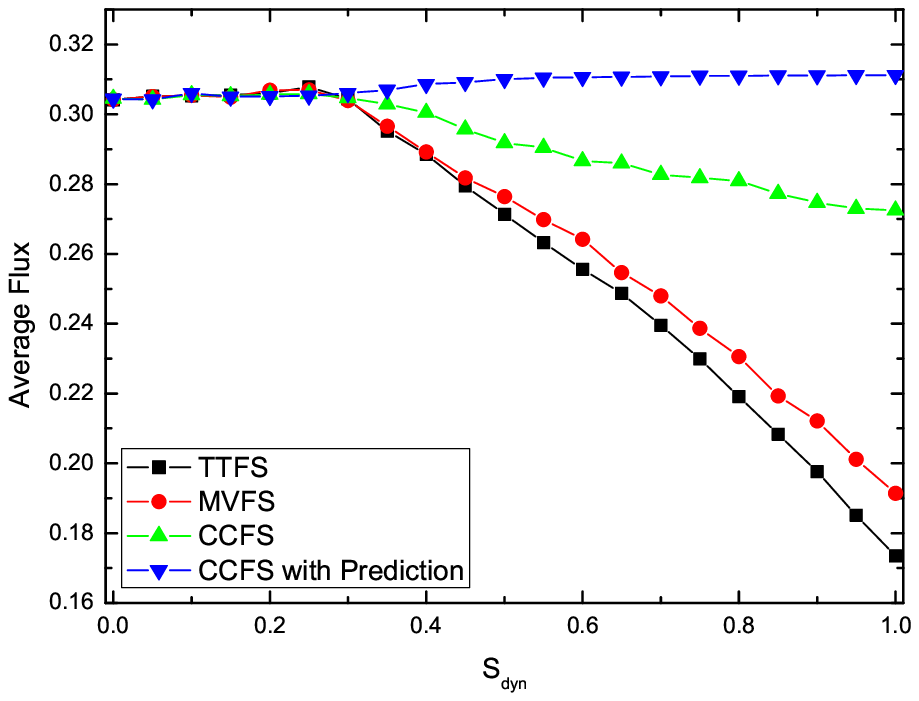}
\caption{\label{fig6} (Color online) Average flux by performing
different strategies \emph{vs} $\emph{S}_{dyn}$ in the three-route
system; \emph{L} is fixed to be 2000, \emph{p} is fixed to be 0.25
and $\emph{T}_{p}$ is fixed to be 260.}
\end{figure}
\begin{figure}
\centering
\includegraphics[scale=1.2]{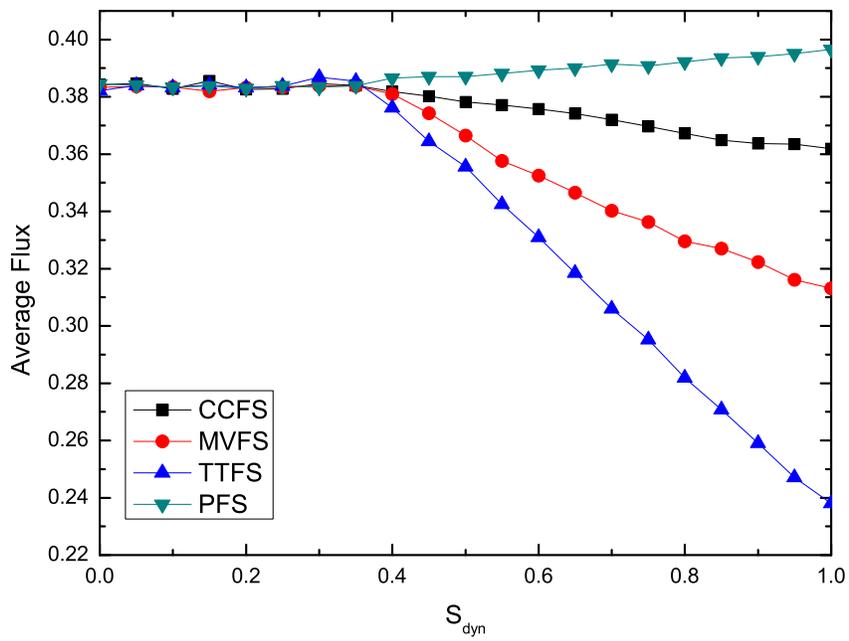}
\caption{\label{fig7} (Color online) Average flux by performing
different strategies \emph{vs} $\emph{S}_{dyn}$ in two-route system;
\emph{L} is fixed to be 2000, \emph{p} is fixed to be 0.25 and
$\emph{T}_{p}$ is fixed to be 60.}
\end{figure}

\vskip 10mm
\section{CONCLUSION}
\par
We obtain the simulation results of applying four different feedback
strategies, i.e., TTFS, MVFS, CCFS and PFS on a three-route scenario
all with respect to flux, number of vehicles, speed, average flux
versus $\emph{T}_{p}$ and average flux versus $\emph{S}_{dyn}$. The
results indicate that PFS has more advantages than the other three
strategies in the three-route system which has only one entrance and
one exit. We also find out that it will take much more time to reach
the stable state in the three-route system than that in the
two-route system. In contrast with the other three feedback
strategies, PFS can significantly improve the road condition,
including increasing vehicle number and flux, reducing oscillation,
and that average flux enhances with increase of $\emph{S}_{dyn}$.
And it can be understood because PFS can eliminate the lag effect.
The numerical simulations demonstrate that the prediction time
($\emph{T}_{p}$) plays a very important role in improving the road
situation.

\par
We also do the simulation of average flux versus $\emph{T}_{p}$ on a
four-route scenario (see Fig.8) which is obtained by 90000
iterations excluding the initial 5000 time steps. We can see that in
Fig.8 the prediction time ($\emph{T}_{p}$) corresponding to the
highest value of the average flux is about 1020 time steps which are
much longer than those in the two-route and three-route systems.
Here ,we can make a reasonable assumption that one car passes the
route with the average speed $v_{mean}$ $\sim$ 2, then the time it
passes the route is $\emph{T}_{pass}$ $\sim$ 1000 because of the
total length of one route \emph{L}=2000. If $\emph{T}_{pass}$ and
$\emph{T}_{p}$ are of the same order of magnitude($\emph{T}_{pass}$
$\sim$ $\emph{T}_{p}$), then the prediction feedback strategy will
become invalid because the car at the entrance of the route at first
will leave the traffic system after $\emph{T}_{pass}$. So we come to
the conclusion that the prediction feedback strategy is appropriate
in multi-route systems when the length of the route is long enough
to ensure $\emph{T}_{pass}$ $>$ $\emph{T}_{p}$.
\begin{figure}
\centering
\includegraphics[scale=1.0]{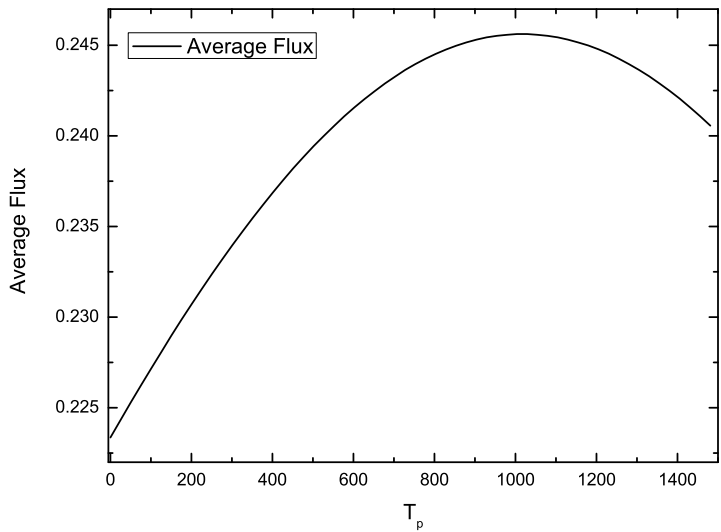}
\caption{\label{fig8} Average flux \emph{vs} prediction
time($\emph{T}_{p}$) in a four-route model. The parameters are
\emph{L}=2000, \emph{p}=0.25, and $\emph{S}_{dyn}$=0.5.}
\end{figure}

\vskip 10mm
\par
\noindent{\large\bf Acknowledgments:}
\par
C.-F Dong would like to thank Dr. Nan Liu at the University of
Chicago and the reviewers for some helpful comments while we were
preparing the manuscript.
\par
This work has been partially supported by the National Basic
Research Program of China (973 Program No. 2006CB705500), the
National Natural Science Foundation of China (Grant Nos. 60744003,
10635040,10532060), the Specialized Research Fund for the Doctoral
Program of Higher Education of China (Grant No. 20060358065) and
National Science Fund for Fostering Talents in Basic Science
(J0630319).

\end{document}